\documentclass[12pt]{article}
\usepackage{psfig}
\begin{document}

\begin{center}
\LARGE{\bf Globular Clusters as a Test for Gravity in the Weak 
Acceleration Regime}\\
\bigskip
\normalsize
{\bf Riccardo Scarpa, Gianni Marconi, and Roberto Gilmozzi}\\

European Southern Observatory \\
E-mail: rscarpa@eso.org, gmarconi@eso.org, rgilmozz@eso.org
\end{center}

\begin{center} {\bf abstract} \end{center}

Non-baryonic Dark Matter (DM) appears in galaxies and other cosmic
structures when and only when the acceleration of gravity, as computed
considering only baryons, goes below a well defined value $a_0=1.2
\times 10^{-8}$ cm s$^{-2}$. This fact is extremely important and
suggestive of the possibility of a breakdown of Newton's law of
gravity (or inertia) below $a_0$, an acceleration smaller than the
smallest probed in the solar system.  It is therefore important to
verify whether Newton's law of gravity holds in this regime of
accelerations. In order to do this, one has to study the dynamics of
objects that do not contain significant amounts of DM and therefore
should follow Newton's prediction for whatever small
accelerations.  Globular clusters are believed, even by strong
supporters of DM, to contain negligible amounts of DM and therefore
are ideal for testing Newtonian dynamics in the low acceleration
limit.  Here, we discuss  the status of an ongoing program
aimed to do this test. Compared to other studies of globular clsuters,
the novelty is that we trace the velocity dispersion profile of globular
clusters far enough from the center to probe gravitational
accelerations well below $a_0$. In all three clusters studied so far
the velocity dispersion is found to remain constant at large radii
rather than follow the Keplerian falloff. On average, the flattening
occurs at the radius where the cluster internal acceleration of
gravity is $1.8\pm 0.4 \times 10^{-8}$ cm s$^{-2}$, fully consistent
with MOND predictions.

\begin{center} {\bf Introduction} \end{center}
We describe here the status of an ongoing observational program aimed
to test the validity of Newtonian dynamics in the low acceleration
limit.  The idea for this project sparked from the consideration that
the typical gravitational accelerations governing the dynamics of
cosmic structures are much smaller than the accelerations probed in
our laboratories or in the solar system. Thus, any time Newton's law
of gravity is applied to galaxies, for instance to infer the existence of
non-baryonic dark matter (hereafter DM), its validity is extrapolated
by several orders of magnitude. Interestingly, unanimous agreement has
been reached (e.g., \cite{binney04}) on the fact that deviations from
Newtonian dynamics are $always$ observed when and only when the
gravitational acceleration falls below a fix value, $\sim 10^{-8}$ cm
s$^{-2}$, as computed considering only baryons. 
This systematic behavior suggests we may be facing a
breakdown of Newton's law rather than the effects of a still
undetected medium. Note that this
acceleration is smaller than the acceleration produced by Mercury on
Pluto, thus even using orbit perturbation theory we really have no way to 
know whether Newtons law of gravity is valid in this regime of accelerations.
 This fact is also at the foundation of a
particular modification of Newtonian dynamics known as MOND
(\cite{milgrom83}), which postulates a breakdown of Newton's law of
gravity (or inertia) below $a_0=1.2\times 10^{-8}$ cm s$^{-2}$
(\cite{begeman91}). MOND successfully describes the properties of a
considerable number of cosmic structures without invoking DM
(\cite{mcgaugh98}; \cite{sanders02}; a short review of MOND is also
given elsewhere in this book).  

Unfortunately, in the realm of
galaxies MOND and DM provides alternative though indistinguishable
descriptions of the data, making it difficult to decide in favor of
one of the two options.

Irrespectively of the validity of MOND, to discriminate between a
failure of Newton's law and DM one as to perform an experiment known
{\it a priori} to be free from the effects of DM. For instance, if
deviations from Newtonian dynamics were to be observed in the
laboratory these will have to be ascribed to a breakdown of Newtonian
dynamics rather than to the effects of DM.  Interestingly, there is
general agreement that, if any, the effects of DM on globular clusters
is dynamically negligible, making them perfect for testing Newtonian
dynamics down to arbitrarily small accelerations.  In previous
papers \cite{scarpa03A} \cite{scarpa03B} \cite{scarpa04} we 
presented the results of a study of the
dynamical properties of the globular clusters $\omega$ Centauri,
M15, and NGC 6171 in which accelerations below $a_0$ were probed. 
It was shown (Figure 1) that
as soon as the acceleration reaches $a_0$ the velocity dispersion
remains constant instead of following the Keplerian falloff.  
This result is similar to what is
observed in galaxies and explained invoking DM, something we can not
do in the present case.

%%%%%%%%%%%%%%%%%%%%%%%%%%%%%%%%%%%%%%%%%%%%
%% Sample figure:
%%
%% The option [height=...] scales the picture to the given height,
%% without it it would be printed at its nominal size
%%%%%%%%%%%%%%%%%%%%%%%%%%%%%%%%%%%%%%%%%%%%

\begin{figure}
\centering
\hbox{\hspace{-0.5cm}
\psfig{file=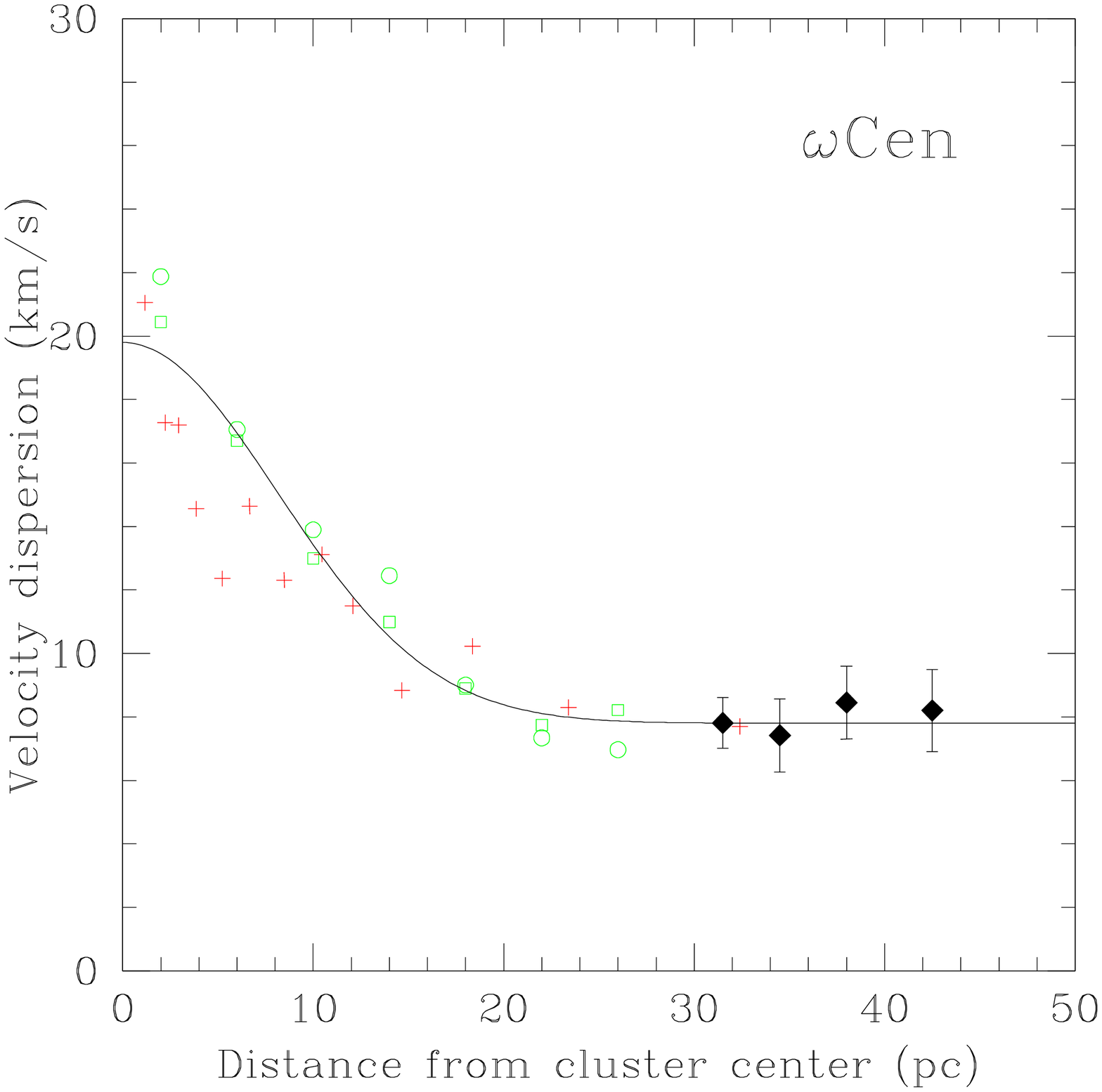,width=0.35\linewidth}\psfig{file=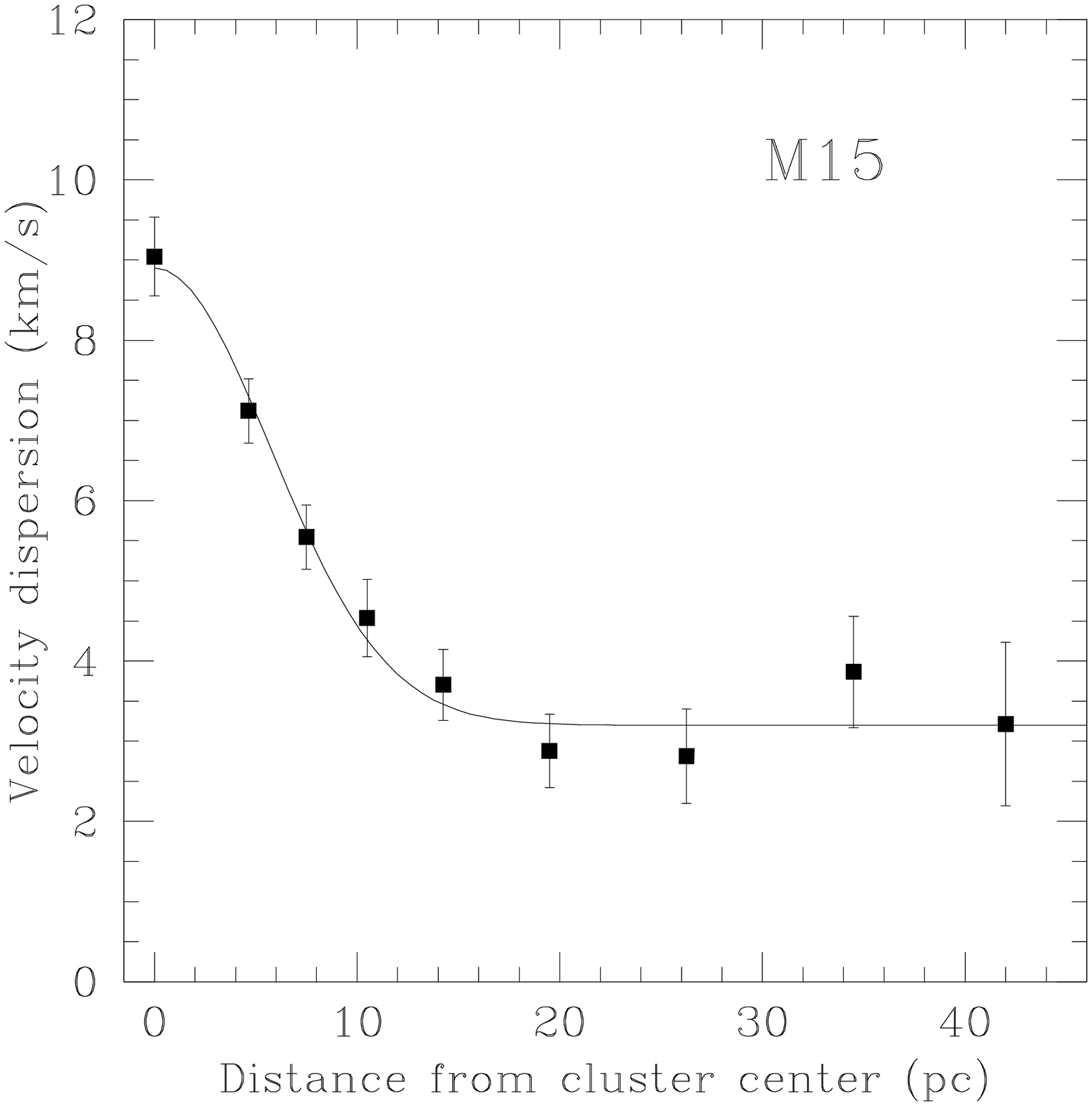,width=0.35\linewidth}\psfig{file=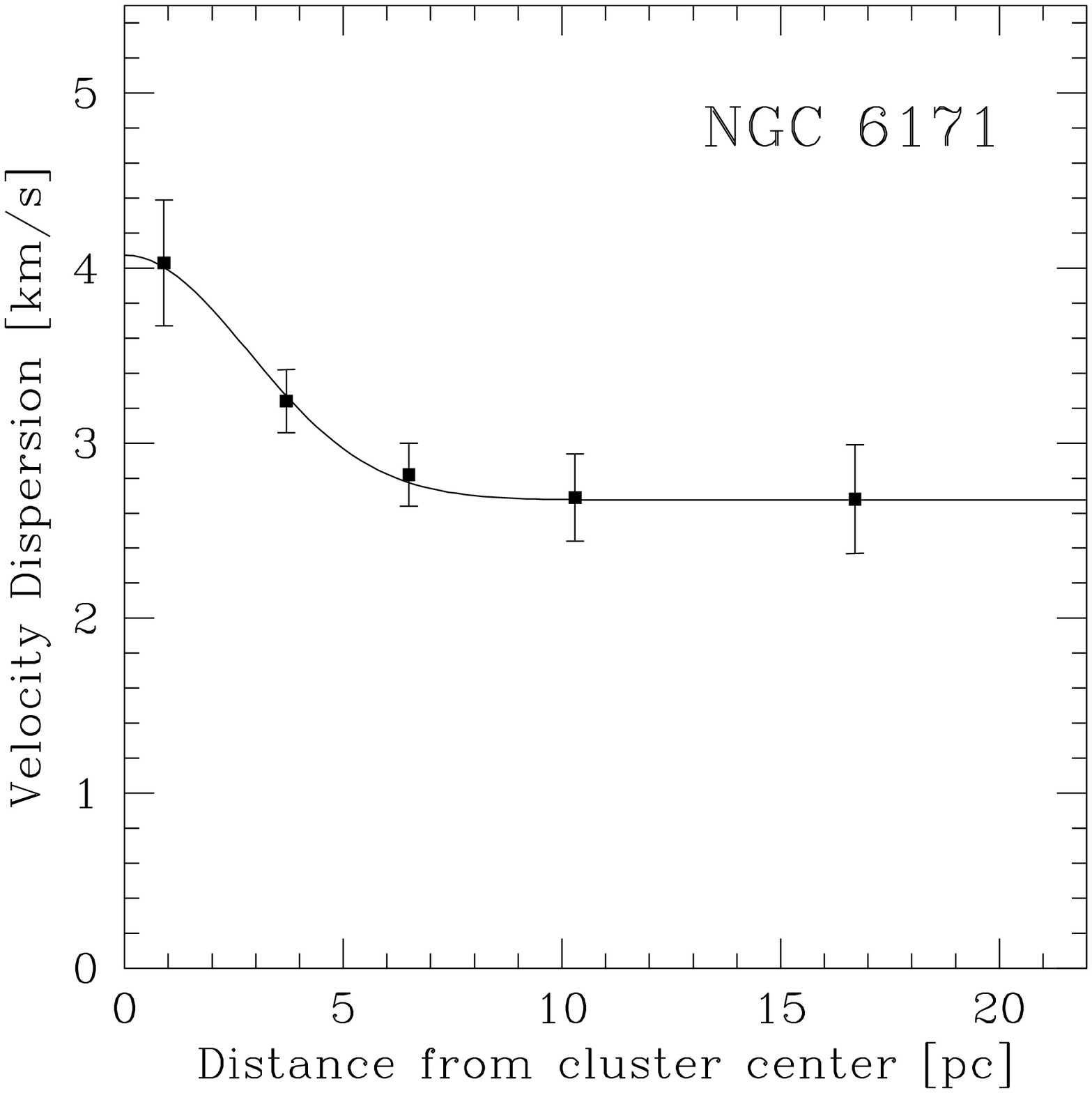,width=0.35\linewidth}
}
\caption{\label{omegacen} {\bf Left:} The velocity dispersion profile
of $\omega$ Centauri (as presented in \cite{scarpa03A}) flattens out
at R$=27\pm 3$ pc, where $a = 2.1\pm 0.5 \times 10^{-8}$ cm s$^{-2}$.
{\bf Center:} Dispersion profile for M15 as derived from data by
\cite{drukier98}. The flattening occurs at R$=18\pm 3$ pc, equivalent
to $a = 1.7\pm 0.6 \times 10^{-8}$ cm s$^{-2}$.  {\bf Right:} Velocity
dispersion profile for NGC 6171 as derived combining our VLT data with
data from \cite{piatek94}.  The profile remains flat within
uncertainties outward of $8\pm1.5$ pc, where
$a=1.4^{+0.7}_{-0.4}\times 10^{-8}$ cm s$^{-2}$.  In all panels, the
solid line is a fit obtained using a Gaussian plus a constant, meant to
better show the flattening of the profile.}
\end{figure}

\begin{center} {\bf Program Status} \end{center}

Assuming a
mass-to-light ratio of one in solar units, in these three clusters the
flattening of the dispersion profile occurs for very similar values of
the internal acceleration of gravity, with an average value of $a =
1.8\pm 0.4 \times 10^{-8}$, fully consistent with the MOND prediction
$a_0$. Considering each of these clusters have a different dynamical history,
tidal heating can not explain this similar behavior. Thus,
existing data support the idea that Newtonian dynamics may not be
applicable for accelerations below $a_0$.
Given the relevance of this claim, this needs confirmation and
generalization. At the time of writing new data have been collected,
at the European Southern Observatory VLT telescope, for the globular
clusters 47 Tuc, NGC 288, and NGC 7099. In total,
approximately 400 stars were observed in 47 Tuc, 240 in NGC 288 and
NGC 7099, with dispersion R$\sim 25000$ so to have an accuracy of few 
hundreds meters per second in the radial velocity. In all cases,
stars were selected based on their position in the H-R diagram. In
particular, stars close to the main sequence turn off were
selected. Even though in principle all targets should be cluster
members, we expect to find a significant fraction of
non-members. Nevertheless, even assuming a 30\% success rate, these
new data should bring to 6 the number of clusters with well studied
velocity dispersion profiles probing accelerations below $a_0$.  In
the case all six will confirm that the velocity dispersion remains
sensibly constant at large radii starting at $\sim a_0$, 
then it would be increasingly
difficult to claim this is due to some conventional effect, like tidal
heating.
 
Of the clusters observed so far, the most important is perhaps NGC
6171.  This is a compact cluster located 6.4 kpc from the sun and only
3.3 kpc from the galactic center (\cite{harris96}). The external
gravitational field due to the Milky Way acting on this cluster is
above $a_0$ for any reasonable value of the Milky Way mass.  According
to Milgrom (\cite{milgrom83}), an object should show MOND effects only
when the total field is below $a_0$. Therefore, we should not observe
a flattening of the dispersion profile in this case, contrary to what
was found.  The last two points of the dispersion profile shown in
Figure 1 are, however, based on 23 and 20 stars,
respectively. Considering the low Galactic latitude of the cluster
contamination may be important here.  Thus we are not in the position
of drawing a firm conclusion about whether MOND effects are seen only
when the total field is below $a_0$.  We therefore have collected data
for an additional $\sim 120$ stars around NGC 6171 to improve the statistical
significance of the result for this cluster.  If it will be possible
to confirm the flattening of the profile, then it would be clear that
only the internal filed is relevant. If this is the case, then it
would become possible to test for deviations from Newtonian dynamics
within the boundaries of the solar system and, in particular, on
earth. With a modern version of the Cavendish experiment it should be
possible to probe the relevant regime of accelerations and detect any
possible deviation from Newtonian dynamics.

%\endinput


\small

\begin{thebibliography}{99}
\bibitem{binney04} Binney J. 2004, in {\it Dark Matter in Galaxies}, 
	ASP conference series, IAUS 220, 3
\bibitem{milgrom83} Milgrom, M. 1983, ApJ 270, 365
\bibitem{mcgaugh98} McGaugh, S., \& de Block, W. J. G. 1998 ApJ 499, 66
\bibitem{sanders02} Sanders, R. H., \& McGaugh, S. S. 2002, ARA\&A 40, 263
\bibitem{begeman91} Begeman, K. G., Broeils, A.H., \& 
	Sanders, R. H. 1991, MNRAS 249, 523
\bibitem{scarpa03A} Scarpa, R., Marconi, G., \& Gilmozzi R. 2003A, A\&AL 405, 15
\bibitem{scarpa03B} ----- 2003B, in {\it Dark Matter in Galaxies}, 
	ASP conference series, IAUS 220, 9
\bibitem{scarpa04} ----- 2004, in ``Baryons in Dark Matter Halos'', 
  Ed. R. Dettmar, U Klein, and P. Salucci. Published by SISSA, 
  Proceedings of Science, http://pos.sissa.it, p. 55.
\bibitem{harris96} Harris, W.E. 1996, AJ, 112, 1487
\bibitem{piatek94} Piatek S. et al 1994, AJ 107, 1397
\bibitem{drukier98} Drukier, G.A., Slavin, S.D., Cohn, H.N. et al 1998, ApJ 115, 708
\end{thebibliography}
\end{document}